\documentclass[twocolumn,prb]{revtex4}
\usepackage{natbib}
\usepackage{graphicx}
\usepackage{subfigure} %% este é o pacote correto para os acentos
\usepackage[latin1]{inputenc}
\usepackage{amsmath}
\usepackage{setspace}
\usepackage{amssymb}
\usepackage{color}

\begin{document}

\title{Glauber Dynamics: An Approach in a Simple Physical System}
\author{Vilarbo da Silva Junior}
\email{vilarbos@unisinos.br}
\affiliation{Centro de Ciências Exatas e Tecnológicas, Universidade do Vale do Rio dos Sinos, Caixa Postal 275, 93022-000 São Leopoldo RS, Brazil}
\author{Alexsandro M. Carvalho}
\email{alexsandromc@unisinos.br}
\affiliation{Centro de Ciências Exatas e Tecnológicas, Universidade do Vale do Rio dos Sinos, Caixa Postal 275, 93022-000 São Leopoldo RS, Brazil}
%\keywords{Spin glass, quantum spherical model, static approximate, exact solution}

\begin{abstract}
In this paper, we investigate a special class of stochastic Markov processes, known as Glauber dynamics. Markov processes are importance, for example, in the study of complex systems. For this, we present the basic theory of Glauber dynamics and its application to a simple physical model. The content of this work was designed in such a way that the reader unfamiliar with the Glauber dynamics, finds here an introductory material with details and example.

\end{abstract}
\maketitle

\section{Introduction}
\label{sec_introd}

Probability theory studies the random phenomena and quantifies their probabilities of occurrence. In principle,
when we observe a sequence of chance experiments, all of the past outcomes could
influence our predictions for the next experiment. For example, the prediction of grades of a student in a sequence of exams in a course.

In 1906, Andrei Andreyevich Markov~\cite{markov1} studied an important type of random process (Markov process). In this process, the outcome of a given experiment can affect the outcome of the next experiment. In other words, the past is conditionally independent of the future given the present state of the process. When presenting the model, Markov did not bother with the applications. In fact, his intention was to show that the large numbers law is valid even if the random variables are dependent. Nowadays, there are numerous applications of which we mention: Biological phenomena~\cite{gibson}, Social science~\cite{stander}, Electrical engineering~\cite{kjersti} among others. We emphasize that physics is one of the areas of knowledge that often uses Markov process. For example, Ehrenfest model~\cite{Ehrenfest} for the diffusion and Glauber dynamics~\cite{Glauber} for the Ising model.

In this paper, we present a special class of Markov processes known as Glauber dynamic~\cite{Glauber,Daniel}. This topic is extremely important since it is the theoretical basis for the Metropolis Algorithm (or simulated annealing~\cite{metropolis}), successfully used to treat and understand problems in physics of complex systems~\cite{Newman,Barabasi}.

The organization this article is as follows: In Section~\ref{sec_premilenary}, we focus in mathematical background (stochastic process and statistical physics). The generalities of the Glauber dynamics are presented in section~\ref{sec_glauberdim}. As an application, we present in section~\ref{secexample1}, the explicit implementation of the Glauber dynamics on the model of localized magnetic nuclei. A physical interpretation of the model is made in section~\ref{sec_resultdisct}. In the sequence, we dedicate the section~\ref{conclusions} to the final considerations.

\section{Preliminary Concepts}
\label{sec_premilenary}

\subsection{Stochastic Process}
\label{subsec_MathBack}

A \textit{continuous time Markov process} on a finite or countable state space $S=\{s_{0},s_{1},\ldots\}$ is a 	
sequence of random variable $X_{0},X_{t_{1}},\ldots$ taking values in $S$, with the property that, for all $t\geq 0$ and $s_{0},s_{1},\ldots, s_{n}, s\in S$ we have 
\begin{equation}
\mathbb{P}(X_{0}=s_{0}|X_{t_{n}}=s_{n})=\mathbb{P}(X_{0}=s_{0},\ldots,X_{t_{n}}=s_{n})
\end{equation}
whenever $t> t_{n}> \cdots > t_{1}> 0$ and $\mathbb{P}(X_{0}=s_{0}|X_{t_{n}}=s_{n})>0$. Here $\mathbb{P}(A|B)=\mathbb{P}(A,B)/\mathbb{P}(B)$ denotes the conditional probability of occur $A$ given that occurred $B$ and $\mathbb{P}(A,B)$ is the probability that occur $A$ and $B$ simultaneously. Thus, given the state of the process at any set of times prior to time $t$, the distribution
of the process at time $t$ depends only on the process at the most recent
time prior to time $t$. This notion is exactly analogous to the Markov property for a
discrete-time process~\cite{norris}.

If the time is discrete, the probability $\mathbb{P}$ associate to the Markov process is completely determined by a time dependent stochastic matrix $P(t)$ (transition probability matrix) and a stochastic vector $\mu$ (initial distribution). When $S$ has $n$ states it follows that $P(t)$ is a $n\times n$ matrix and its elements are denoted as $p_{ij}(t)=\mathbb{P}(X_{t}=j|X_{0}=i)$, i.e, $p_{ij}(t)$ is the transition probability from $i$ to $j$ in time $t$. Furthermore, the $n$ coordinates of $\mu$ are $\mu_{i}=\mathbb{P}(X_{0}=i)$ represents the probability of finding the system in state $i$ initially. A initial distribution $\mu$ are called \textit{invariant} or \textit{stationary} from $P(t)$ if satisfies $\mu P(t)=\mu$ for all $t\geq 0$. This fact indicates the equilibrium state of the system. If there is a invariant distribution $\mu$ from $P(t)$, them we have $\lim_{t\rightarrow \infty}P(t)=M$ where all the lines of $M$ are $\mu$.

Now, if time is continuous, the dynamics of $P(t)$ is find as solution of the initial value problem (Kolmogorov Equation~\cite{norris})
\begin{eqnarray}
\label{eq_2GlauberEdu}
    \frac{d}{dt}P(t)&=&Q P(t),  \\
    P(0)&=&I \nonumber,
\end{eqnarray}
where $I$ is the identity matrix and the $Q$ matrix is called $Q$-\textit{matrix}. The elements of Q, $q_{ij}$, must satisfy the following conditions:
\begin{eqnarray}
\label{eq_3GlauberEdu}
0\leq -q_{ii}< \infty, \;\; &\forall & \, i\\
\label{eq_4GlauberEdu}
q_{ij}>0\;\; &\forall & \, i\neq j\\
\sum_{j\in S}q_{ij}=0\;\; &\forall & \, i
\label{eq_5GlauberEdu}
\end{eqnarray}
The $Q$-matrix is also called matrix row sum zero, which complies with its last property. Each off-diagonal entry $q_{ij}$ we shall interpret as the rate going from $i$ to $j$ and the diagonal elements are chosen in general as $q_{ii}=-\sum_{j\neq i}q_{ij}$. For end, the explicit form of transition probability matrix is $P(t)=e^{tQ}$ (solution of differential equation~(\ref{eq_2GlauberEdu})), where the exponential matrix is $e^{tQ}=\sum_{k=0}^{\infty}t^{k}Q^{k}/k!$. Thus, the stochastic vector $\mu(t)$ that describes the probability of finding the system in their states on time $t$ is $\mu(t)=\mu e^{tQ}$, where $\mu$ is the initial distribution. The solution $P(t) = e^{tQ}$ shows how basic the generator matrix $Q$
is to the properties of a continuous time Markov chain. For example, if $\nu$ is a probability vector and satisfies $\nu Q=0$ means $\nu$ is a stationary distribution of Markov process\cite{norris}.

Some stochastic processes have the property that, when the direction of time is reversed the behavior of the process remains the same. This class of stochastic process is known as \textit{reversible stochastic process}. We say that a continuous time Markov process $\{X_{t}\}_{t\geq 0}$ is \textit{reversible} with respect to initial distribution $\mu$ if, for all $n\geq 1$, $s_{0},s_{1},\ldots,s_{n}\in S$ and $t_{n}>\cdots>t_{1}>0$ occur $\mathbb{P}(X_{0}=s_{0},X_{t_{1}}=s_{1}\ldots,X_{t_{n}}=s_{n})=\mathbb{P}(X_{t_{n}}=s_{0},X_{t_{n}-t_{n-1}}=s_{1}\ldots,X_{0}=s_{n})$. Intuitively, if we take a movie of such a process and,  then, to run the movie backwards the process result will be statistically indistinguishable from the original process. There is a condition for reversibility that can be easily checked, called \textit{detailed balance condition}. In more detail, this condition is obtained when Markov process is reversible with respect to initial distribution $\mu$, i.e.,
\begin{equation}
\mu_{i}q_{ij}=\mu_{j}q_{ij},
\label{eq_6GlauberEdu}
\end{equation}
for all states $i$ and $j$, where $q_{ij}$ denotes the entry of Q-matrix of $P(t)$ and $\mu_{i}$ the coordinates of $\mu$.
A first consequence is that if $\mu$ and $Q$ satisfies the detailed balance condition then $\mu$ is a stationary distribution for $P(t)$~\cite{norris}. This condition has a very clear intuitive meaning: in equilibrium, we must have the same number of transitions in both directions ($i\rightarrow j$ and  $j\rightarrow i$).

\subsection{Statistical Physics}
\label{subsec_MecStatBack}

Here, we present some concepts of equilibrium statistical mechanics. For a complete treatment, we suggest the references: Huang~\cite{huang} and Reif~\cite{reif}.

\textit{Simple systems} are macroscopically homogeneous, isotropic, discharged, chemically inert and sufficiently large. Often, a simple system is called \textit{pure fluid}. A \textit{composite system} is constituted by a set of simple systems separated by \textit{constraints}. Constraints are optimal partitions that can be restrictive to certain variables. The main types of constraints are: \textit{adiabatic}, \textit{fixed} and \textit{impermeable}.

In relation the equilibrium thermodynamics is important to know its postulates, which are:

\begin{itemize}

\item \textit{First Postulate}: The microscopic state of a pure fluid is completely characterized by the internal energy $U$, volume $V$ and number of particles $N$.

\item \textit{Second Postulate}: There is a function of all extensive parameters of a composite system named entropy $S(U_{1},V_{1},N_{1},\ldots,U_{n},V_{n},N_{n})$, which is defined for all equilibrium states. On removal of an inner constraints, the extensive parameters assume values which maximize the entropy.

\item \textit{Third Postulate}: The entropy of a composite system is additive on each of its components. Entropy is a continuous, differentiable and monotonically increasing function.

\item \textit{The fundamental postulate of statistical mechanics}: In a closed statistical system with fixed energy, all accessible microscopic states are equally likely.

\end{itemize}
\begin{figure}[t!]
\centerline{
\includegraphics[width=0.3\textwidth]{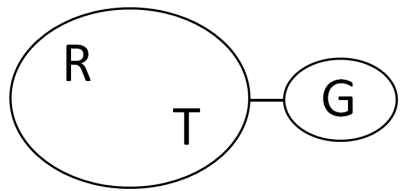}}
\caption{Simple system $G$ in contact with a thermal reservoir $R$ with temperature $T$.}
\label{fig1}
\end{figure}

Let us consider a simple system $G$ in contact with a thermal reservoir $R$ with temperature $T$, by means of a diathermic constraint fixed and impermeable (see Fig.~\ref{fig1}), where $R$ is very large compared to $G$. If the composite system $G+R$ is isolated with total energy $E_{0}$, then the probability distribution \begin{equation}
\label{eq:GibbsState}
\mu(T)=(\mu_{1}(T),\mu_{2}(T),\ldots)
\end{equation}
is characterized by
\begin{equation}
\mu_{i}(T)=\frac{e^{-\frac{E_{i}}{T}}}{Z(T)},
\label{eq_7GlauberEdu}
\end{equation}
where $\mu_{i}(T)$ is the probability of finding the system $G$ in particular microscopic state $i$ with energy $E_{i}$ and temperature $T$ (we choose the Boltzmann constant $k_{B}=1$, for convenience). The normalization constant $Z(T)=\sum_{j}e^{-E_{j}/T}$ is called \textit{partition function} of the system. Furthermore, the distribution $\mu(T)$ is known as a \textit{Gibbs states}. Thus the \textit{canonical ensemble} consists in a set microscopic states $i$ accessible to the system $G$ in contact with a thermal reservoir $R$ and temperature $T$, with probability distribution given by Eq.~(\ref{eq_7GlauberEdu}) (Gibbs distribution). Clearly, there is a energy fluctuation in the canonical ensemble. Using the Gibbs distribution, we obtain that the average energy of system $G$ (using $\beta=1/T$) is
\begin{equation}
\left<E\right>=-\frac{\partial }{\partial \beta}\log{(Z(\beta))}=\sum_{j}\mu_{j}(\beta)E_{j}
\label{eq_8GlauberEdu}
\end{equation}
and its variance can be write as
\begin{equation}
\sigma^{2}(E)=\left<E^{2}\right>-\left<E\right>^{2}=\frac{\partial^{2} }{\partial \beta^{2}}\log{(Z(\beta))},
\label{eq_9GlauberEdu}
\end{equation}
where $E_{j}$ is the energy in a particular microscopic state $j$.

In summary, based on the Gibbs distribution, equilibrium statistical mechanics indicates that states with lower energy are more likely than those with higher energy.

\section{Glauber Dynamics}
\label{sec_glauberdim}

We saw in the previous section, the Gibbs state (\ref{eq:GibbsState}) is the equilibrium state of the system. Thus, a relevant question is: how the system evolves from the initial state to the Gibbs state?  Note that the equilibrium state, in the context of the Markov process, corresponds to stationary distribution. Thus, we forward the answer to the question as the solution of Kolgomorov equation (\ref{eq_2GlauberEdu}). However, for this purpose, we need to know the behavior of the Q-matrix. An alternative to the shape of the Q-matrix is the Glauber dynamics~\cite{Glauber}.
 
In order to build a Glauber dynamics is necessary to know the single particle energy function $E_{i}$ and their accessible microscopic states (states space) $S=\{i\}$. Thus, we write explicitly the partition function $Z(T)$ and Gibbs states $\mu_{i}(T)$. To describe the time dependent probabilities matrix $P(t)=e^{tQ(T)}$ which is reversible with respect to the Gibbs state, we need to propose a $Q(T)$-matrix that satisfies the detailed balance condition (\ref{eq_6GlauberEdu}) with Gibbs state $\mu(T)$ (for each fixed temperature $T$). There are many other possibilities~\cite{sinai}, and the optimal choice is often dictated by special features of the situation under consideration. However, for our purposes, the one which will serve us best is the one
whose $Q(T)$-matrix is given by~\cite{Daniel}
\begin{eqnarray}
\begin{cases} q_{ij}(T)=e^{-\frac{1}{T}(E_{j}-E_{i})} &\mbox{if}\,\, E_{j} > E_{i} \\
q_{ij}(T)=1 & \mbox{if}\,\, E_{j}\leq E_{i}\\
q_{ii}(T)=-\sum_{j\neq i}q_{ij}(T),
\end{cases}
\label{eq_10GlauberEdu}
\end{eqnarray}
where $E_{i}$ is the single particle energy at particular microscopic state $i$.

In the App.~\ref{canddetbalanc1}, we show that above matrix satisfies the detailed balance condition with the Gibbs states. Therefore, the time dependent probability matrix generated by Eq.~(\ref{eq_10GlauberEdu}) is a Glauber dynamics for the Gibbs states.

\section{Example: Localized Magnetic Nuclei}
\label{secexample1}

\begin{figure}[t!]
\centerline{
\includegraphics[width=0.48\textwidth]{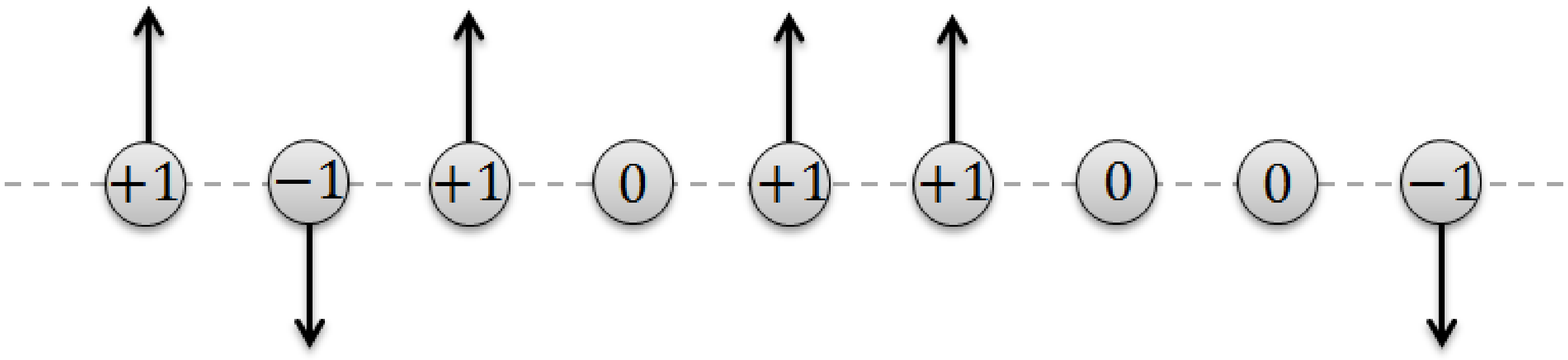}}
\caption{Picture of a magnetic nuclei chain. The up arrow means $\sigma=+1$, no arrow means $\sigma=0$ and down arrow $\sigma=-1$.}
\label{fig2}
\end{figure}

The nuclei of certain solids~\cite{sessoli} have integer spin. According to quantum theory~\cite{sakurai}, each nuclei can have three quantum spin states (with $\sigma=+1,0$ or $-1$). This quantum number measures the projection of the nuclear spin along the axis of the crystalline solid. As the charge distribution is not spherically symmetric, nuclei energy depends on the spin orientation relative to the local electric field. In Fig.~\ref{fig2}, we show a possible configuration of this magnetic nucleons.

Thus, nuclei in states $\sigma= \pm 1$ and $\sigma=0$ have energy, respectively,  $D>0$ and zero.
Therefore, its energy function is given by
\begin{equation}
E_{\sigma}=D\sigma^{2},
\label{eq_11GlauberEdu}
\end{equation}
where $D>0$ is electric field intensity and the microscopic states (quantum states) are characterized by random variables (spins) $\sigma\in S=\{+1,0,-1\}$. So, for each fixed temperature $T$, the partition function is written as
\begin{equation}
Z(T)=\sum_{\sigma\in \{+1,0,-1\}}e^{-\frac{1}{T}D\sigma^{2}}=1+2e^{-\frac{1}{T}D}
\label{eq_12GlauberEdu}
\end{equation}
and the Gibbs states coordinates $\mu_{\sigma}(T)$ (\ref{eq_7GlauberEdu}) are given by
\begin{equation}
\mu_{\sigma}(T)=\frac{e^{-\frac{1}{T}D\sigma^{2}}}{1+2e^{-\frac{1}{T}D}}.
\label{eq_13GlauberEdu}
\end{equation}

In order to explicit the $Q(T)$-matrix for this physical system, we apply the energy function (\ref{eq_11GlauberEdu}) at (\ref{eq_10GlauberEdu}). This results (see App.~\ref{qmatrixcalculo})
\begin{equation}
Q(T)=\left(
       \begin{array}{ccc}
         -2 & 1 & 1 \\
         e^{-\frac{1}{T}D} & -2e^{-\frac{1}{T}D} & e^{-\frac{1}{T}D} \\
         1 & 1 & -2 \\
       \end{array}
     \right).
\label{eq_14GlauberEdu}
\end{equation}

To find $P(t)=e^{tQ(T)}$, we need to diagonalize the $Q(T)$-matrix~\cite{lay}. For this, we must present a invertible matrix $B$ such as $Q(T)B=BD_{3}$ (equivalently $Q(T)=BD_{3}B^{-1}$), where $B$´s columns is composite by eigenvector of $Q(T)$, $B^{-1}$ its inverse and $D_{3}$ is a diagonal matrix ($3\times 3$) formed by eigenvalues of $Q(T)$. In this present case, the characteristic polynomial is
\begin{eqnarray}
p(\lambda)&=&\det{(Q(T)-\lambda I)} \nonumber \\
&=&-e^{-\frac{1}{T}D}\lambda (3+\lambda)(2+e^{\frac{1}{T}D}(1+\lambda)).
\label{eq_15GlauberEdu}
\end{eqnarray}
The eigenvalues $\lambda$'s are solution of characteristic equation $p(\lambda)=0$, i.e, 
\begin{equation}
\lambda_{1}=-3,\qquad \lambda_{2}=0,\qquad \lambda_{3}=-Z(T),
\label{eq_16GlauberEdu}
\end{equation}
where $Z(T)$ is the partition function (\ref{eq_12GlauberEdu}).  Consequently, its associated eigenvectors are
\begin{equation}
v_{1}=\left(
        \begin{array}{c}
          -1 \\
          0 \\
          1 \\
        \end{array}
      \right),\:v_{2}=\left(
        \begin{array}{c}
          1 \\
          1 \\
          1 \\
        \end{array}
      \right),\: v_{3}=\left(
        \begin{array}{c}
          1 \\
          -2e^{\frac{1}{T}D} \\
          1 \\
        \end{array}
      \right).
\label{eq_17GlauberEdu}
\end{equation}
We conclude that $Q(T)$-matrix admits a decomposition $Q(T)=BD_{3}B^{-1}$ (see App.~\ref{diagppelements}). Then, $P(t)=Be^{tD_{3}}B^{-1}$ is responsible for describing the dynamics of transition probabilities between spin. More explicit, $P(t)$ is
\begin{equation}
P(t)=\left(
       \begin{array}{ccc}
         p_{+1+1}(t) & p_{+10}(t) & p_{+1-1}(t) \\
         p_{0+1}(t) & p_{00}(t) & p_{0-1}(t) \\
         p_{-1+1}(t) & p_{-10}(t) & p_{-1-1}(t) \\
       \end{array}
     \right)
\label{eq_18GlauberEdu}
\end{equation}
where
\begin{eqnarray*}
p_{+1+1}(t)&=&p_{-1-1}(t)=\frac{e^{-3t}}{2}+\frac{e^{-\frac{1}{T}D}}{Z(T)}+\frac{e^{-Z(T)t}}{2Z(T)},\\
p_{0+1}(t)&=&p_{0-1}(t)=\frac{e^{-\frac{1}{T}D}}{Z(T)}-\frac{e^{-\frac{1}{T}D-Z(T)t}}{Z(T)},\\
p_{-1+1}(t)&=&p_{+1-1}(t)=-\frac{e^{-3t}}{2}+\frac{e^{-\frac{1}{T}D}}{Z(T)}+\frac{e^{-Z(T)t}}{2Z(T)},\\
p_{+10}(t)&=&p_{-10}(t)=\frac{1}{Z(T)}-\frac{e^{-Z(T)t}}{Z(T)},\\
p_{00}(t)&=& \frac{1}{Z(T)}+\frac{2e^{-\frac{1}{T}D-Z(T)t}}{Z(T)}.
\end{eqnarray*}
Here, $p_{\sigma,\widetilde{\sigma}}(t)=\mathbb{P}(X_{t}=\widetilde{\sigma}|X_{0}=\sigma)$ indicates the transition probability from spin state $\sigma$ to $\widetilde{\sigma}$ in time $t$. For example, $p_{+10}(t)$ is the probability from $\sigma = +1$ to $\sigma=0$ after time $t$.

It is relatively easy to prove that $P(t)$ satisfies the detailed balance condition with Gibbs state $\mu(T)$ (\ref{eq_13GlauberEdu}). Therefore, the stochastic Markov process $\{X_{t}\}_{t\geq 0}$ where the random variable $X_{t}$ denotes the quantum spin state of each located nucleus at time $t$ (i.e. $X_{t}=\sigma$) is a Glauber dynamics.

\section{Results and Discussion}
\label{sec_resultdisct}

\begin{figure}[t!]
\centerline{
\includegraphics[width=0.48\textwidth]{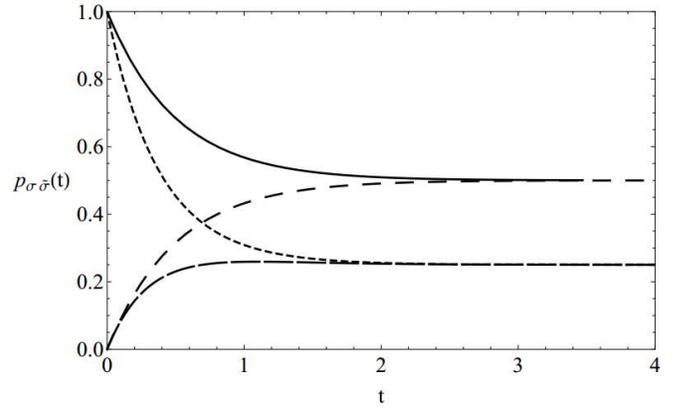}}
\caption{Time dependence for some transition probabilities $p_{\sigma\widetilde{\sigma}}(t)$ for $T=1$ and $D=\log{(2)}$. The solid line corresponds to $p_{00}(t)$, short dashed line $p_{+1+1}(t)$, medium dashed line $p_{+10}(t)$ and long dashed line $p_{+1-1}(t)$.}
\label{fig3}
\end{figure}

Given the $P(t)$ elements and an initial quantum state, we can follow the dynamics of the transition probability between quantum spin states. For example, consider initially the state $\sigma=+1$. Note that we have $p_{+1+1}(t)\geq p_{+1-1}(t)$ for all $t$. This means that since the nuclei is in the quantum spin state $\sigma=+1$ is more likely that it remains in such a state that it ``flip'' to the quantum spin state $\sigma=-1$. In the Fig.~\ref{fig3}, we present the dynamics of some transition probabilities. For our choice of parameters $T=1$ and $D=\log{(2)}$ we have $\lim_{t\rightarrow \infty}\mu(T)(t)=\mu(T)=(1/4\,\,\,1/2\,\,\,1/4)$. Therefore, in a sample of $N$ nuclei, on average $N/2$ occupy the quantum spin state $\sigma=0$, $N/4$ occupy $\sigma=+1$ and $N/4$ occupy $\sigma=-1$. In this figure, we see the convergence (exponential) this limit as well as the consistency in their values.

Still on the limit $t\rightarrow \infty$, we have
\begin{equation}
\lim_{t\rightarrow \infty}P(t)=\left(
                                 \begin{array}{ccc}
                                   \mu_{+1}(T) & \mu_{0}(T) & \mu_{-1}(T) \\
                                   \mu_{+1}(T) & \mu_{0}(T) & \mu_{-1}(T) \\
                                   \mu_{+1}(T) & \mu_{0}(T) & \mu_{-1}(T) \\
                                 \end{array}
                               \right)
\label{eq_19GlauberEdu}
\end{equation}
where $\mu_{\sigma}(T)$ are the coordinates of Gibbs state $\mu(T)$. This result is consistent with that shown in the Sec.~\ref{subsec_MathBack}, i.e, $\mu(T)=(\mu_{+1}(T) \,\, \mu_{0}(T) \,\, \mu_{-1}(T))$ is the unique equilibrium state (invariant distribution) for $P(t)$. Then, whatever the initial spin states distribution $\nu=(\nu_{+1}\,\, \nu_{0}\,\,\nu_{-1})$, we always have $\lim_{t\rightarrow \infty}\mu(T)(t)=\lim_{t\rightarrow \infty}\nu P(t)=\mu(T)$.
This result, partly, justifies freedom of choice in the initial distribution of spin states in Monte Carlo simulations (Metropolis algorithm~\cite{binder1}).

Note by Eq.~(\ref{eq_13GlauberEdu}) that $\mu_{0}(T)\geq \mu_{+1}(T)=\mu_{-1}(T)$. Thus, we can conclude that after a long time, most of the nuclei are occupying the quantum spin state $\sigma=0$. This occupation number is due to the fact that the quantum spin state $\sigma=0$ is less energetic than the other two states ($E_{0}=0$ and $E_{+1}=E_{-1}=D>0$). This is in accordance with a general physical principle that any physical system tends to occupy the lower energy state. Additionally, note that if $T\rightarrow \infty$, we prove that $\lim_{T\rightarrow \infty}\mu_{0}(T)=\lim_{T\rightarrow \infty}\mu_{+1}(T)=\lim_{T\rightarrow \infty}\mu_{-1}(T)=1/3$. Therefore, for high temperature the three quantum spin states are equally likely (same occupation number). In the opposite limit, when $T\rightarrow 0$, $\lim_{T\rightarrow 0}\mu_{0}(T)=1$ and $\lim_{T\rightarrow 0}\mu_{+1}(T)=\lim_{T\rightarrow 0}\mu_{-1}(T)=0$. This indicates that for low temperatures, all the nuclei tend to occupy the quantum spin state of lower energy $\sigma=0$ (\textit{ground state}). Another important verification is the average number of spin $\left<\sigma\right>$, given by
$\left<\sigma\right>=\sum_{\sigma\in S}\sigma \mu_{\sigma}(T)=0 $
where $S=\{+1,0,-1\}$. If we chose a nuclei randomly, is more likely to be found in the quantum spin state $\sigma=0$. The quantity $m=\left<\sigma\right>$ is called the \textit{magnetization} of system~\cite{reif}.

\section{Conclusions}
\label{conclusions}

This paper presents a review about continuous time stochastic Markov processes which are reversible with respect to Gibbs State,  called Glauber dynamics.

The main result of our exposition is contained in Sec.~\ref{secexample1}. We use the theory developed in the preceding sections applying them in a very simple physical model. This example, contains the necessary ingredients to illustrate richness of the method.

We show that, after a long time interval, distribution of quantum spin states are given by the state Gibbs.  This state is a equilibrium state for the Glauber dynamics. So any initial distribution of spin state will relax in $\mu(T)$. This,  ,in turn, justifying the free choice of the initial distribution of spin states in computational simulations via Monte Carlo method.

For a fixed temperature, we verify that the quantum spin state $\sigma = 0$ is the one with the highest number of occupants (most likely). This is consistent with what is expected physically since $\sigma = 0$ is the lowest energy state. This fact indicates that average value of the random spin variable was zero and, consequently, a zero magnetization for a sample of this kind of spins. 

Finally, for high temperatures, the nuclei are uniformly distributed in the quantum spin states ($1/3$ for each). On the other hand, for low temperatures, all the nuclei tend to occupy a quantum spin state $\sigma = 0$.

%Finally, we have developed an introductory material on Glauber dynamics containing a didactic and methodology structure in their exposure, allowing a graduate student in physics can find here a starting point on the topics covered.
%

\section{Appendices}
\label{apedicies}

\subsection{Proof of the Detailed Balance Condition}
\label{canddetbalanc1}

In order to show the equality (\ref{eq_6GlauberEdu}) for $Q(T)$-matrix (\ref{eq_10GlauberEdu}) and Gibbs state (\ref{eq_7GlauberEdu}) let us assume, without loss of generality, that $E_{j}>E_{i}$. So,
\begin{eqnarray*}
\mu_{i}(T)q_{ij}(T)&=&\frac{1}{Z(T)}e^{-\frac{E_{i}}{T}}e^{\frac{1}{T}(E_{j}-E_{i})}\\
&=&\frac{1}{Z(T)}e^{-\frac{E_{i}}{T}}e^{-\frac{E_{j}}{T}}e^{\frac{E_{i}}{T}}\\
&=&\frac{1}{Z(T)}e^{-\frac{E_{j}}{T}}e^{-\frac{E_{i}}{T}+\frac{E_{i}}{T}}\\
&=&\frac{1}{Z(T)}e^{-\frac{E_{j}}{T}}1=\mu_{j}(T)q_{ji}(T),
\end{eqnarray*}
because as we are assuming $E_{j}> E_{i}$ it follows that $E_{i}< E_{j}$ then by (\ref{eq_10GlauberEdu}) $q_{ji}(T) = 1$. Therefore, $Q(T)$ and $\mu(T)$ satisfy the detailed balance condition and hence $P(t)=e^{tQ(T)}$ generated by $Q(T)$ is a Glauber dynamics.

\subsection{$Q(T)$-matrix Calculations}
\label{qmatrixcalculo}

In this appendix, we show in detail the entry of $Q(T)$-matrix. As the state space $S=\{+1,0,-1\}$ has tree states our $Q(T)$-matrix is $3\times 3$ and its elements $q_{\sigma,\widetilde{\sigma}}(T)$, evaluated as prediction in Eq.~(\ref{eq_10GlauberEdu}) with Eq.~(\ref{eq_11GlauberEdu}), are given by:
\begin{eqnarray*}
q_{+10}(T)&=&e^{-\frac{1}{T}(E_{0}-E_{+1})}=1,\\
q_{+1-1}(T)&=&e^{-\frac{1}{T}(E_{-1}-E_{+1})}=e^{-\frac{1}{T}(D-D)}=1,\\
q_{+1+1}(T)&=&-q_{+10}(T)-q_{+1-1}(T)=-2,\\
q_{0+1}(T)&=&e^{-\frac{1}{T}(E_{+1}-E_{0})}=e^{-\frac{1}{T}(D-0)}=e^{-\frac{1}{T}D},\\
q_{0-1}(T)&=&e^{-\frac{1}{T}(E_{-1}-E_{0})}=e^{-\frac{1}{T}(D-0)}=e^{-\frac{1}{T}D},\\
q_{00}(T)&=&-q_{0+1}(T)-q_{0-1}(T)=-2e^{-\frac{1}{T}D},\\
q_{-1+1}(T)&=&e^{-\frac{1}{T}(E_{+1}-E_{+1})}=e^{-\frac{1}{T}(D-D)}=1,\\
q_{-10}(T)&=&e^{-\frac{1}{T}(E_{0}-E_{-1})}=1,\\
q_{-1-1}(T)&=&-q_{-1+1}(T)-q_{-10}(T)=-2.
\end{eqnarray*}
because $E_{\pm 1}>E_{0}$.

\subsection{Explicit Elements: $B$, $D_{3}$ and $B^{-1}$ }
\label{diagppelements}

In Sec.~(\ref{secexample1}), we discus a decomposition $Q(T)=BD_{3}B^{-1}$. Explicitly,
\[
B=\left(
  \begin{array}{ccc}
    -1 & 1 & 1 \\
    0 & 1 & -2e^{-\frac{1}{T}D} \\
    1 & 1 & 1 \\
  \end{array}
\right), \qquad
D_{3}=\left(
    \begin{array}{ccc}
      -3 & 0 & 0 \\
      0 & 0 & 0 \\
      0 & 0 & -Z(T) \\
    \end{array}
  \right)
\]
and
\[
B^{-1}=\left(
         \begin{array}{ccc}
           -\frac{1}{2} & 0 & \frac{1}{2} \\
           \frac{e^{-\frac{1}{T}D}}{Z(T)} & \frac{1}{Z(T)} & \frac{e^{-\frac{1}{T}D}}{Z(T)} \\
           \frac{1}{2Z(T)} & -\frac{1}{Z(T)} & \frac{1}{2Z(T)} \\
         \end{array}
       \right).
\]

%\begin{acknowledgments}
%We Thank Professor Rogério Ricardo Steffenon by frequent encouragement.
%\end{acknowledgments}

\bibliographystyle{phcpc} 

\end{document}